\documentclass[pdflatex,sn-basic,iicol]{sn-jnl}

\newcommand\apjs{ApJS}               
\newcommand\aap{A\&A}                
\newcommand\ao{Appl. Opt.}           
\newcommand\apjl{ApJ}                

\jyear{2022}%

\theoremstyle{thmstyleone}%
\theoremstyle{thmstyletwo}%

\theoremstyle{thmstylethree}%

\begin{document}

\title[UV Spectropolarimetry of Colliding Wind Binaries ]{UV Spectropolarimetry with {\it Polstar}: Massive Star Binary Colliding Winds}

\author*[]{\fnm{Nicole }\sur{St-Louis$^{*,1}$}}\email{nicole.st-louis@umontreal.ca}

\author*[]{\fnm{Ken }\sur{Gayley$^{2}$}}

\author*[]{\fnm{D. John }\sur{Hillier$^{3}$}}

\author*[]{\fnm{Richard }\sur{Ignace$^{4}$}}

\author*[]{\fnm{C. E. }\sur{Jones$^{5}$}}

\author*[]{\fnm{Alexandre }\sur{David-Uraz$^{6,7}$}}

\author*[]{\fnm{Noel D. }\sur{Richardson$^{8}$}}

\author*[]{\fnm{Jorick S. }\sur{Vink$^{9}$}}

\author*[]{\fnm{Geraldine J. }\sur{Peters$^{10}$}}

\author*[]{\fnm{Jennifer L. }\sur{Hoffman$^{11}$}}

\author*[]{\fnm{Ya\"el }\sur{Naz\'e$^{12}$}}

\author*[]{\fnm{Heloise }\sur{Stevance$^{13}$}} 

\author*[]{\fnm{Tomer }\sur{Shenar$^{14}$}} 

\author*[]{\fnm{Andrew G. }\sur{Fullard$^{15}$}} 

\author*[]{\fnm{Jamie R.} \sur{Lomax$^{16}$}}

\author*[]{\fnm{Paul A. }\sur{Scowen$^{17}$}}








\abstract{The winds of massive stars are important for  their direct impact on the interstellar medium, and for their influence on the final state of a star prior to it exploding as a supernova.
However, the dynamics of these winds is understood primarily via their illumination from a
single central source.
The Doppler shift seen in resonance lines is a useful tool for inferring these dynamics, but 
the mapping from that Doppler shift to the radial
distance from the source is ambiguous.
Binary systems can reduce this ambiguity by providing
a second light source at a known radius in the wind, seen from orbitally
modulated directions.
From the nature of the collision between the winds, a massive companion also provides unique additional information about wind momentum fluxes.
Since massive stars are strong ultraviolet (UV) sources, and UV resonance line opacity in
the wind is strong, UV instruments with a high resolution spectroscopic capability are essential
for extracting this dynamical information.
Polarimetric capability also helps to further resolve ambiguities in aspects of the
wind geometry that are not axisymmetric about the line of sight,
because of its unique access to scattering direction information.
We review how the proposed MIDEX-scale mission \textit{Polstar} can use UV spectropolarimetric
observations to critically constrain the physics of colliding winds, and hence radiatively-driven winds in general.  We propose a sample of 20 binary targets, capitalizing on this unique combination of illumination by companion starlight,
and collision with a companion wind, to probe wind attributes over a range in wind strengths.
Of particular interest is the hypothesis that the radial distribution of the wind acceleration
is altered significantly, when the radiative transfer within the winds becomes optically thick
to resonance scattering in multiple overlapping UV lines.}




\maketitle

\section{Introduction}\label{sec1}

Despite comprising the rarest stellar mass group, massive stars ($>$ 8 M$_\odot$) are amongst the most important originators of elements in the Universe because they synthesize and distribute heavy elements, especially oxygen through aluminum, when they explode as supernova.  They also form and evolve much faster than lower-mass stars, thereby influencing the formation and composition of all other stars and their planets. The drastically stronger stellar wind of a massive star, when compared to the solar wind, also enriches the interstellar medium, and can lead to significant changes in the star's mass over its lifetime. This affects its evolutionary path in the H-R diagram, its type of supernova, whether it becomes a gamma-ray burster, and the compact remnant it produces, along with any gravitational wave signature it might create
if it is a close binary.

Indeed,
most massive stars spend a large fraction of their lives in binary systems with other massive stars, and more than half are thought to engage in mass exchange with a close companion \citep{Sana2012}. 
Tracking these evolutionary effects are the topic of two other \textit{Polstar} proposal objectives \citep[][]{jonestc, peterstc};
this paper relates to the objective that entails using the proximity of
the two massive stars as a unique probe of the nature of their wind dynamics.

\subsection{Wind Acceleration and Stratification}

Over the past few decades our understanding of radiatively-driven hot, massive-star winds has greatly improved.
But despite this
observational and theoretical progress, fundamental details of radiatively driven winds, such as the velocity structure, still elude us. The so-called {\em $\beta-law$}, although extremely convenient to describe the general density structure of massive-star winds, does not arise from a solid physical description of the winds; it rests on an unlikely simplification that the wind opacity does not
undergo significant change as the wind temperature and density drop with radius, and as the
radiation field shifts from primarily diffusive to primarily radially streaming. 
Attempts to better describe the radial acceleration profile of hot-star winds have been made over the years, such as adjusting the value of the $\beta$ exponent, or in the case of WR stars, adopting a typical value of that exponent in the inner wind, while using in the outer wind a higher
value describing a second extended acceleration regime \citep{Hillier1999,Grafener2005}. More recently, \citet{Sander2020} presented hydrodynamically consistent models of classical WR stars and find that a traditional {\em $\beta-law$} is inadequate for mass-loss rates above 10$^{-6}$ M$_{\odot}$ yr$^{-1}$ \citep[see also][]{Grafener2005}.  Instead, a plateau in the velocity law develops due to the decrease of the radiative acceleration between the two opacity bumps (hot and cool).

\begin{figure*}
\centering
\includegraphics[width=\textwidth]{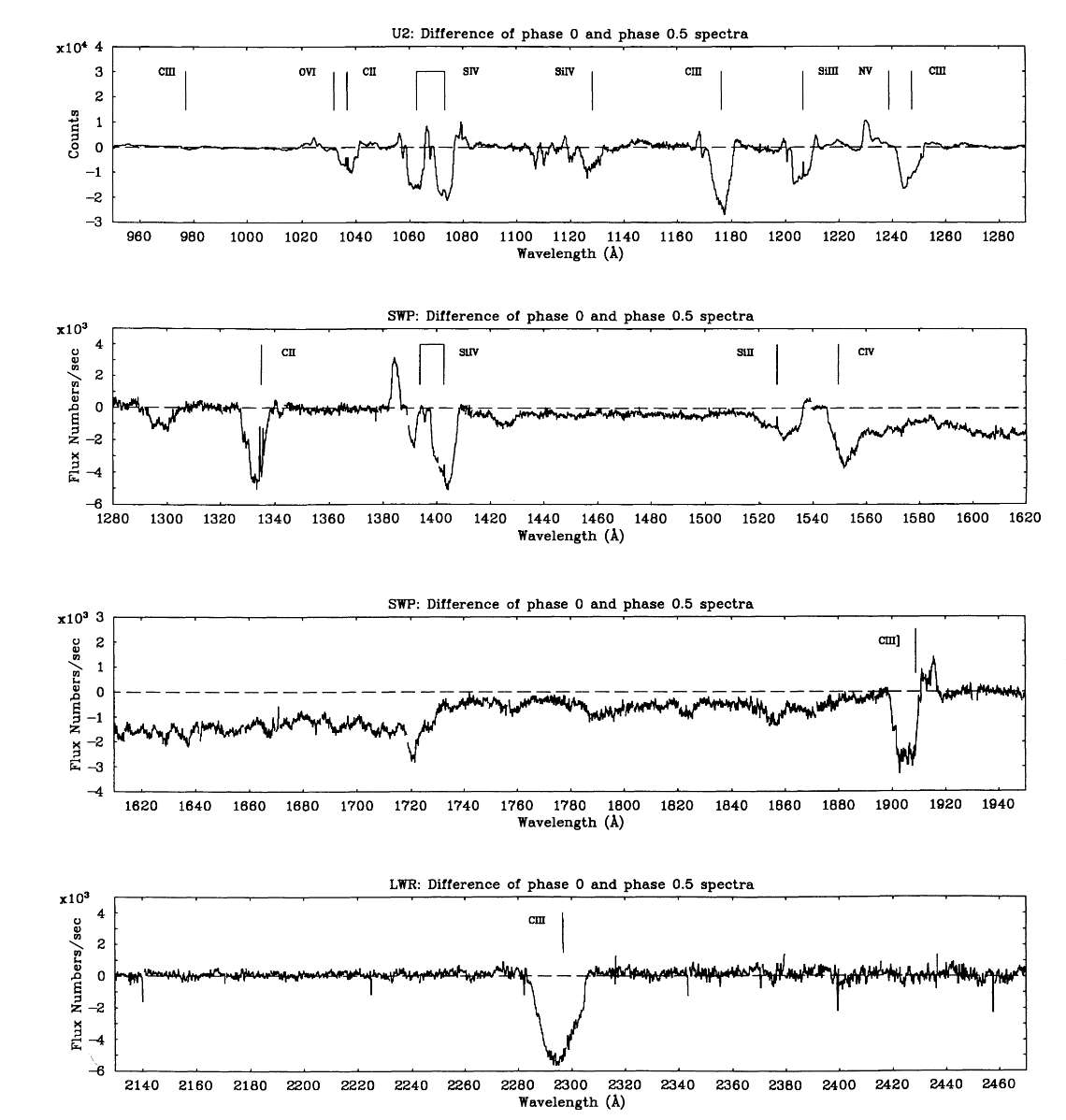}
\caption{Eclipse spectrum of $\gamma$ Velorum. Figure 2 from \citet{1993ApJ...415..298S}. Reproduced with permission.}
\label{figGVdif}
\end{figure*}

The radial dependence of the wind speed has
important consequences, as it controls the steady-state density structure, and creates the diagnostically
significant mapping from observable Doppler shift to radial location.
In optically thick outflows such as those of WR stars, the detailed wind structure is even more important, 
because the hydrostatic surface of the star is hidden from view, leaving the wind as
the only layer of the star that is directly accessible to observation.  
Dense winds also experience additional opacity changes as the temperature drops significantly 
and the radiation field shifts from being highly diffusive in deep layers to more free streaming in outer layers. Therefore, we wish to test the hypothesis that the different character of radiative transport
and opacity gradients in denser winds, compared to less dense winds, changes the nature of the
acceleration as a function of radius.
This paper describes how this test can be carried out in a special sample of colliding-wind
binary systems.



{\centering
\begin{figure}[t]
\includegraphics[width=8.0cm]{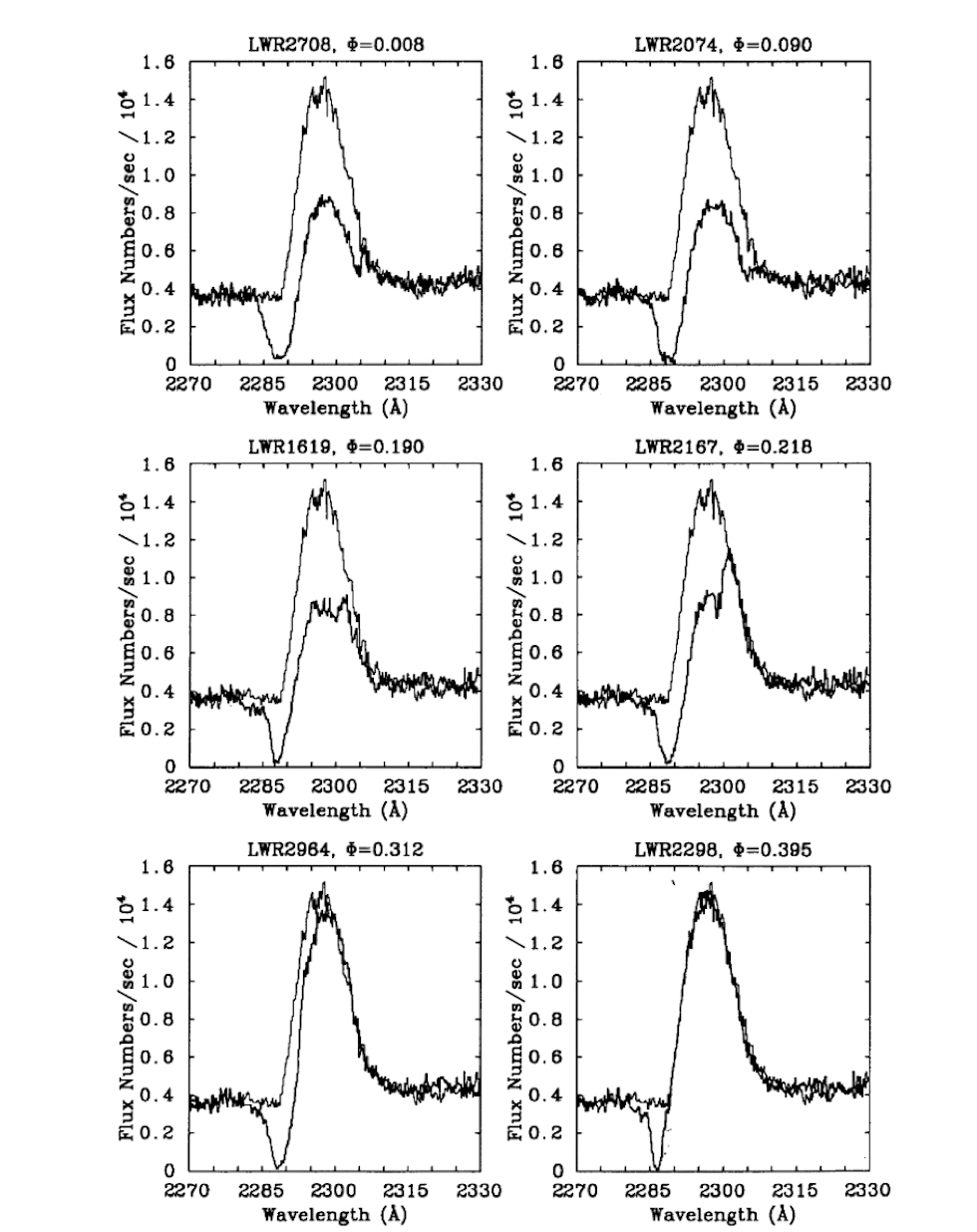}
\caption{ Variations in the C {\sc iii} $\lambda$ 2297 profile with orbital phase. The thin line is the LWR 1316 ($\Phi = 0.534$), and the thick line is the spectrum and phase indicated at the top of each graph. Figure 4 from \citet{1993ApJ...415..298S}. Reproduced with permission.}
\label{figGV}
\end{figure}
}

\subsection{Making Use of Massive Stars in Binaries}

Massive stars in binaries offer a unique opportunity to improve our understanding of radiatively driven winds. We can use the light from one star to probe the wind of the other star, allowing us to study the
structure of its wind, which provides information on its density, the velocity law, 
the acceleration zone, and the ionization structure. 
This method has long
been successfully used in the past to investigate, for example, the nature of WR stars
\citep{1993ApJ...415..298S}.

An even older version of this approach that focused on light curves \citep{Cherepashchuk1984} 
from the WN5 + O6 V eclipsing binary V444 Cygni, covering the wavelength range from $\sim$ 2400\,\AA\  to 3.5\,$\mu m$, demonstrated 
that the WN5 star was hot (90,000 K), compact ($r(\tau_{\scriptstyle}=1)=2.9\,R_\odot$), and had an outflowing atmosphere in which the temperature decreased smoothly with height. 
This helped resolve a long-standing controversy in the nature of Wolf-Rayet stars. 
Additional studies have investigated the stratification in the atmosphere \citep[e.g.,][]{Eaton1985,Brown1986}, and revealed, for example that Fe\,{\sc v} in the wind of the WR star contributed to the atmospheric eclipse signal in the UV.

\cite{Koenigsberger1985} used observations of 6 WR+O systems to show the extended WR atmosphere eclipsed the O stars, and used it to provide constraints on the atmosphere. A further analysis of four binaries by \cite{Koenigsberger1990} showed that the Fe\,{\sc iv}/Fe\,{\sc v} ratio decreased with radius, and that the wind was still being accelerated at 14\,$R_\odot$. \citet{Auer1994} completed a detailed study and showed how line profile variations could be used to investigate the structure of the winds.
Thus, select binary systems have a history of generating breakthrough studies of hot stars, a
spirit that is extended into the present in this paper.

\subsection{The Diagnostic Potential of Wind Eclipses}

 For a given binary configuration, spectroscopic variations are caused by the absorption of the light from one star by the atoms in the wind of the other.
 Previous UV spectroscopy has shown that absorption occurs mainly in low-excitation and resonance transitions of abundant ions \citep[e.g.][]{Koenigsberger1985}. Therefore, these were called {\em selective wind eclipses}. As shown in Figure \ref{figGVdif}, for the WR+O binary $\gamma$ Velorum \citep{St.-Louis1993}, many ions show a symmetric absorption profile spanning positive and negative velocities in the difference spectrum between phases 0 (WR star in front) and 0.5 (O star in front), as expected. However, note that for certain lines, such as the semi-forbidden C\,{\sc iii}] $\lambda$1909 line and the entire forest of Fe\,{\sc iv} lines, only absorption at negative velocities (towards observer) is observed. This unexpected behaviour still remains to be explained.

Examination of the times series for the same star of lines that behave more as
expected has shown that the absorption profile is widest when the O companion is directly behind the WR star
(called phase 0), and becomes narrower as it orbits to a perpendicular direction with the line-of-sight
(phase 0.5).
This is seen in Figure \ref{figGV}) by considering the difference between the line profile at
the phase shown for the well-isolated C~\textsc{iii} $\lambda$ 2297 line, and that line seen
when the O companion is nearly perpendicular to the WR star.
Here there is evidence of wind eclipse when the O star is seen through the WR wind, at both
red and blue Doppler shifts, spanning the range of projected velocities in the WR wind.

The detailed shape of the eclipse profile in the various transitions will be determined by the opacity encountered by the light from the companion star on its way to the observer and therefore on the projected velocity and ion abundance. As the companion moves through its orbit, different parts of the wind are probed as the line-of-sight differs for different configurations.  This allows to probe different regions of the flow as a function of distance from the center of the star. If the two stars have strong winds, each one can act as a probe for the wind of the other. The combination of all resulting eclipse profiles observed at each position of the orbit contains information on the velocity, density and abundance structures of the winds.

Because of the presence of strong resonance lines of several abundant ions (e.g., due to C\,{\sc iv}, 
Si\,{\sc iv}), far ultraviolet spectroscopy is an ideal tool to study massive-star winds and particularly the spectroscopic variations originating from selective wind eclipse. Emission at any given frequency probes a large volume of the wind whereas absorption features primarily probes material along the line of sight to the star. The shape of the continuum light-curve over the binary orbit also allows to probe the column depth of the wind. The opacity of O and WR star winds at optical and UV wavelengths is primarily due to electron opacity.
In WR stars Fe lines in the UV create a pseudo``continuum", and these also contribute significantly to the opacity.

In Figure\, \ref{Fig_rtau} we illustrate the location at which a radial optical depth of 2/3 occurs as a function of wavelength for an O star and a WR star. The minimum radius, at these wavelengths, is set by electron scattering. At other wavelengths we see the influence of strong bound-bound transitions. The Fe forest is particularly prevalent in the WR star. While the actual numerical values are dependent on the
adopted stellar radius, effective temperature and mass-loss rate, the plots do show how different wavelengths have the potential to be used as diagnostic probes in binary star systems.  


{\centering
\begin{figure}
\centering
\includegraphics[width=8.0 cm]{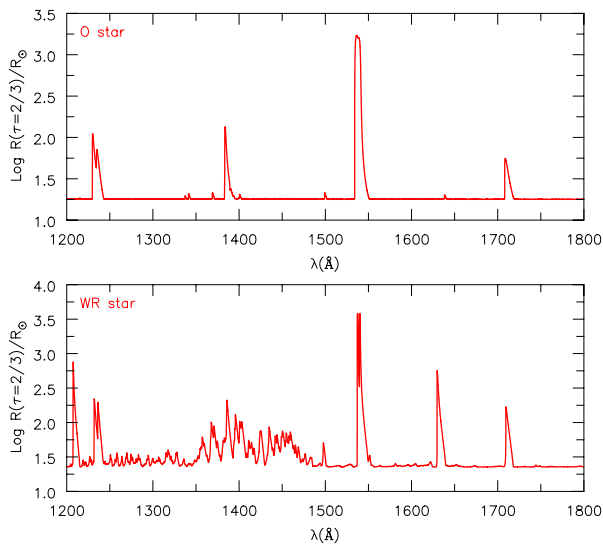}
\caption{Illustration of the location of where a
radial optical depth of $\tau=2/3$ occurs as a
function of wavelength for an O-type and W-R type star.}
\label{Fig_rtau}
\end{figure}
}

\subsection{Colliding Winds in Massive Binaries}

Another aspect of close massive binaries is that the winds of the two stars collide forming a shock-cone structure, generating an asymmetric gas distribution.  Colliding-wind binaries (CWB) can teach us a great deal about the individual stars and their winds because the geometry of the interaction region is strongly dependent on the component winds. Indeed, the details of the geometry and physical characteristics of the shock structure depend on the physical parameters of the individual winds, such as the mass-loss rate and velocity structure, thereby providing a further probe of these parameters. 

Early theories describing these binaries used momentum flux \citep[e.g.][]{Girard1987} and ram-pressure balance \citep[e.g.][]{Kallrath1991}, or hydrodynamic models \citep[e.g.][]{Stevens1992}, until \citet{Canto1996} produced a complete analytical description in the limit of rapid shock cooling (see Section \ref{sec:canto}). Further work in the area has focused on hydrodynamic simulations \citep[][]{Pittard2009,Lamberts2011,Macleod2020} and predictions of line profiles in both optical \citep{Luehrs1997,Ignace2009_pol,Georgiev2004} and X-rays \citep{Rauw2016_FeXXV,Mossoux2021}. The result of these models, combined with a number of phase-resolved observations \citep[e.g.][]{1990MNRAS.243..662W,1999Natur.398..487T, Rauw2001,Sana2001,Rauw2002,Sana2008,Gosset2009,Parkin2009, 2009MNRAS.395.2221W,2010ApJ...709..632K, Lomax2015,Naze2018,2019MNRAS.488.1282W, Callingham2020}, is that we have a good understanding of the expected geometry of colliding winds.

However, collision of the winds will additionally induce complex geometric structures that must be carefully analyzed to be able to extract the desired diagnostics.  In this context, purely spectroscopic information may not suffice to resolve the potential ambiguities in such a complex structure, including the location and density of the bow shock, the inclination of the orbit, and the potential for asphericity in the intrinsic winds. 

Polarimetric observations are thus a vital tool for revealing details of the mass-loss and mass-transfer structures in these binaries. Both continuum emission (arising from the stars) and line emission (arising from winds, CIRs, or other shock regions) may be polarized in a CWB system via scattering from wind clumps \citep{Harries1998,Davies2005}, accretion disks \citep{Hoffman1998}, jets \citep{Fox1998}, bow shocks \citep{Shrestha2018,Shrestha2021}, and other asymmetric distributions of material or velocities within the system \citep{Schulte-Ladbeck1992,Villar-Sbaffi2005,Fullard2020}. Because scattering near the orbital plane tends to produce repeatable effects over many binary cycles, spectropolarimetric monitoring allows us to reconstruct the 3-D shapes of the regions scattering both continuum and line emission (e.g., \citealt{St.-Louis1993,Lomax2015,Fullard2020}).

Given that massive star winds are strongly ionized, it is reasonable to consider Thomson scattering as the dominant scattering method for photons in the winds, which in turn can polarize the observed light. The resulting polarization is sensitive to the geometry of scattering regions but also depends on the chemical composition of the wind and its  stratification as this will strongly affect the number of available scatterers. Therefore modeling polarimetric observations is a powerful tool to inform us about the geometry and density structures of the colliding winds in interacting systems. This information is critical to our understanding of the roles of binary stars in enriching the ISM and producing explosive transients and gravitational-wave sources.

The classic  model approximates the time-varying continuum polarization caused by the illumination of stellar winds in a binary system viewed at an arbitrary inclination angle. In this model, the scattering region is described as an optically thin electron gas. The envelope pattern is assumed to co-rotate with the illumination sources, appropriate for steady-state systems in circular orbit. The illuminators consist of one point source at the center of the scattering region and an additional external point source representing the companion. 

\citet{Brown1982} extended the classic model developed by \citet[][]{Brown1978} (hereafter BME) describing the variation of continuum polarization from the illumination of circumstellar material in a circular binary to consider elliptical orbits, and \citet{Fox1994} further extended the formalism to consider finite illuminators. \citet{Fox1994} showed that occultation is only important  in very close binary systems, where separation is less than 10 times the radius of the primary. \citet{Hoffman2003} quantified the polarization produced by external illumination of a disk in a binary system. However, none of these enhancements to the theory included the effects of colliding winds in the time-dependent polarization results. \citet{St.-Louis1993} and \citet{Kurosawa2002} modeled the polarization variations in the colliding-wind system V444 Cyg.  Also, spectroscopic analysis of wind eclipses in general and for this system in particular have been carried out \citep[][]{1994ApJ...436..859A}, but the results have not been compared across a range of binary parameters to extract the important trends.  \citet{HarriesBablerFox2000} used a Monte Carlo formalism to estimate the polarisation caused by the scattering off dust and electrons in a colliding-wind shock cone if it were present in the long-period binary WR137 and found that for such a wide binary, the level would be very small (0.03\%\,). Notwithstanding these individual efforts,  a general formalism has not yet been produced. 

The \textit{Polstar} team includes several experts in polarimetric modeling who will focus on developing the necessary tools to derive physical results from spectropolarimetric signatures. Polarization simulations of bow shocks, which polarize light both when illuminated by a central source and when the light originates within the shock itself \citep{Shrestha2018,Shrestha2021}, can be incorporated into binary models, as we discuss below. We will also investigate the effects of the wind collision regions and the distributed nature of line emission on the wavelength dependence of polarization. Given that complex continuum and line polarization signals associated with CWBs have been observed in several systems \citep{St.-Louis1993,Lomax2015,Fullard2020} and are likely to occur in the UV as well \citep{Schulte-Ladbeck1992,Hoffman1998}, such modeling efforts are critically important. 

We intend to use \textit{Polstar} to secure extensive UV spectropolarimetric timeseries of key massive binary systems in crucial lines with observations well-distributed around the orbital cycle.  Having multiple diagnostics is important -- different diagnostics probe different regions of the flow and trace different physics. For example, P~Cygni absorption (blue shifted absorption associated with a redshifted emission profile) only samples material along our line of sight. Conversely, the emission line  samples a much large volume. Continuum polarization, to some extent, is influenced by  the whole volume. Optical studies do not have the necessary lines for the studies proposed here. A well-distributed  time series is also crucial, since spectra taken at different times probe distinct orientations
of the binary system. The combination of polarimetry and UV spectroscopy, almost unexplored, offers a unique opportunity to resolve ambiguities left by either approach alone.  For example, the rate of change of the polarization position angle caused by the colliding wind interaction is a diagnostic of the orbital inclination that is complementary to light-curve considerations.

\subsection{Small and Large-Scale Structure in Radiatively-Driven Winds}

Growing observational evidence suggests that
radiatively driven outflows are highly clumped, including the
telltale presence of small sub-peaks observed to move from the center to the edges of strong emission lines in both O stars \citep[e.g.,][]{Eversberg1998} and Wolf-Rayet (WR) stars \cite[e.g.,][]{Lepine1999}.  Inconsistencies in line strengths \citep[e.g.,][]{Hillier2003,Crowther2002,Massa2003,Fullerton2006} and the relative strength of electron scattering wings to their adjacent profiles \citep[e.g.][]{Hillier1991,Hamann1998} also indicate that the winds are clumped. Large-scale structures are also thought to be present in the winds of most, if not all, OB stars, as shown by the presence of NACs and DACs in their ultraviolet spectra \citep[e.g.,][]{Howarth1989,Kaper1996} that are thought to be the observational signature of Corotating Interaction Regions \citep[CIRs,][]{1996ApJ...462..469C, Mullan1986}.

These important structural effects 
\citep[][]{GayleyTC}
but can be further elucidated by the special diagnostics available in colliding-wind binaries.
Small and large-scale wind structures change the density structure of the winds, and this influences the nature of the wind acceleration being probed in this paper.  
Further, winds may be porous both spatially (porosity) and in velocity space \citep[called {\em vorosity};][]{Oskinova2007, Owocki2008}. Spatial porosity affects both line and continuum radiation transfer whereas vorosity, which refers to gaps in velocity space, only effects line transfer. These phenomena lead to uncertainties in derived parameters such as mass-loss rates, luminosities, radii, and abundances. In turn, this limits our ability to place accurate constraints on stellar and galactic evolution.
As will be seen below, two very bright systems ($\gamma$ Velorum and $\delta$ Ori)
that will be targeted for close examination for 
clumping effects are also targeted here by virtue of their binary status, and these systems will
be especially informative when the \textit{Polstar} diagnostics utilized in the two objectives are combined.

\subsection{Mass-loss rates}

Mass-loss rates included in stellar evolution calculations typically use a fitting law with an adjustable parameter \citep[e.g. MESA or Geneva code, ][]{Paxton2011, Meynet2015}. These laws are based on a combination of theoretical and observational studies of mass loss in O stars.  For radiatively-driven winds of hot stars, the theoretical mass-loss rates of \cite{Vink2001} and \cite{Nugis2000} are typically adopted. For more luminous stars, they are in rough agreement (factor of two) with empirically derived mass-loss rates, but for late O stars the rates can differ by an order of magnitude.
Even a factor of two uncertainty is a significant problem for understanding how massive stars evolve
and effect their environment.

Observational mass-loss rates are generally derived using an empirical velocity law, and by necessity one is 
forced to make assumptions regarding the inhomogeneity of the wind \cite[e.g.,][]{Hillier2003,Bouret2012,Sander2012, Shenar2015}. Typically, the wind is assumed to be clumped, with a fraction $f$ occupied by the clumps. Porosity is generally ignored, and the interclump medium is assumed to be void. To first order, empirical modeling derives $\dot M/\sqrt{f}$.  Constraints on $f$ can be placed by using the strength of electron scattering wings in WR stars, and by using P Cygni profiles in O stars. The latter, however, is more uncertain because of the effects of porosity and vorosity.

Current radiative transfer codes can, potentially, derive the velocity law and mass-loss rates from first principles \cite[e.g.,][]{Vink2001,Sander2020,Bjorklund2021,Vink2021}. However such codes make assumptions about clumping, and may use approximate radiation transfer techniques such as the use of the Sobolev approximation \citep{Sobolev1960}. Further, there are still uncertainties about the nature and role of microturbulence at the sonic point
\citep{Hillier2003,Lucy2007}, and the role of inflation, convection and density inhomogeneities in optically thick regions of the flow \citep[e.g.,][]{Ishii1999,Petrovic2006,Cantiello2009,Grafener2012,Grassitelli2016}. Finally there are uncertainties in the atomic data. 

As a consequence of these uncertainties, spectra derived from models constructed on first principle generally do not provide a good fit to observed spectra. At best,  mass-loss rates must be regarded as uncertain to at least a factor of two, and in some cases the uncertainty is much larger. However, a factor of two variation in mass-loss rate can have a strong influence on a star's evolution. This is seen in theoretical calculations \citep[e.g.,][]{Maeder1981}, and is likely illustrated by the different massive star populations in the LMC and Galaxy, as line-driven mass-loss rates depend strongly on metallicity \citep[e.g.][]{Vink2001}.
Accurate mass-loss rates require an understanding of the density structure, which connects to the
velocity structure and the physics of wind driving.
Thus the objective considered in this paper is also relevant to establishing more accurate
mass-loss rates in specially selected systems, 
using the unique diagnostics available in these binary systems.

\section{PolStar Capabilities and Data Products}

Using \textit{Polstar} \citep{ScowenTC}, we intend to probe the details of the wind density of the stars in these systems by using the light from the other star as a probe.  Because the presence of two stars with strong winds in a binary generally leads to the formation of a shock-cone structure, we must also take this into account and actually we intend to use it to further constrain the physical characteristics of the outflows. Finally, companion stars are also not independent light sources -- their radiation field can modify the wind of the other star, and in close binaries one or both winds may not reach terminal speed either because of the collision or because of the companion's radiation field. Radiative braking, the deceleration of one wind by the radiation field of the secondary, will modify the location and structure of the bow shock. The combination of UV spectroscopic time series, polarimetric observations, and spectra modeling will ultimately allow us to constrain the mass-loss rates to unprecedented accuracy.

Phase-dependent observations of systems with elliptical orbits or with different separations will allow us to investigate how the radiation field of one star influences the wind of the second star. Our strategy to address these questions is to study a wide range of wind densities in binary systems with high enough inclination and short enough periods for the companion light to be seen deeply through the primary wind. For the brighter stars, we intend to obtain a full phase coverage of ultraviolet spectropolarimetric observations for each system at the highest possible spectral resolution.  Depending on the brightness of a given target, we can bin the spectral resolution as necessary to secure sufficiently high quality signal-to-noise to identify telltale spectral features and polarization variability, with good phase coverage.

We have identified a sample of 20 massive binary systems for which there is evidence in the literature of the presence of a colliding wind structure, and which can be observed with Polstar. 
These are listed in Table 1.

\begin{table*}[t]
\tiny 
\caption{Target list }
    \centering

        \begin{tabular}{l                         |c|c|c|c|c|c|c}
       
     \hline\hline   Star & HD number & Type & Period (d) & V & UV Flux* &Ch. 1$\dagger$  &Ch.2$\dagger$ \\
        &&&&&&exp time(s)&exp time(s)\\
        \hline
        \hline
&&&&&&&\\
{\bf Ch.1 spec., N=1}&&&&&&&\\
{\bf and polar. (SNR=3000)}&&&&&&&\\
Gamma Velorum & 68273 & WC8+O7.5III-V & 78.05 & 1.83 & 263 & 2000 & 60\\
Delta Ori & 36486 & O9.5II+BO5.III & 5.7 & 2.41 & 147 & 3550& 60\\
29 CMa & 57060 & O7Iaf+O9 & 4.4 & 4.95 & 7.5 & 70200& 60 \\ 
&&&&&&&\\
\hline
&&&&&&&\\
{\bf Ch.1 spec., N=1 (SNR=100)}&&&&&&&\\
{\bf Ch. 2 polar. (SNR=3000)}&&&&&&&\\
Plaskett & 47129 & O7.5I+O6I & 14.4 &  6.06 & 1 & 600 & 1840\\ 
WR22 & 92740 & WN7h+O9III-V &80 & 7.16 & 1.1 & 510&1680\\
   &152248 & O7.5III(f)+O7III(f) & 5.8 & 6.05 & 0.8 & 720 &2300\\
WR79 & 152270 & WC7+O5-8 & 8.89 & 6.95 & 0.5 &1150&3680\\ 
& 149404 & O7.5If+ON9.7I & 9.8 & 5.52 & 0.5 & 1150&3680\\
&&&&&&&\\
\hline
&&&&&&&\\
{\bf Ch.1 spec., N=1.5 (SNR=100)}&&&&&&&\\
{\bf Ch. 2 polar. (SNR=3000)}&&&&&&&\\
WR133 & 190918 & WN5o+O9I & 112.8& 6.7 & 0.35 & 1110&585\\
WR42 & 97142 & WC7+O5-8 & 7.886 & 8.25 & 0.22 & 1800 &930\\
Eta Car(2020) & 93308B & LBV & 2022.7 & 6.21 & 0.2 & 1950&1050\\
WR 25 & 93162 & O2.5If*/WN6+O & 208 & 8.84 &  0.13&4020&1590\\
WR140 & 193793 & WC7pd + O5.5fc & 2895.0  & 6.85 & 0.1 &4020 & 2100\\
& 93205 &O3.5 V((f)) + O8 V &6.0803 &7.75&0.1 & 6300&2100\\
&&&&&&&\\
\hline
&&&&&&&\\
{\bf Ch.1 spec., N=3 (SNR=100)}&&&&&&&\\
{\bf Ch. 2 polar. (SNR=1000)}&&&&&&&\\
WR137 & 192641 & WC7+O9e & 4766 & 8.15 & 0.04 & 5100 & 5100\\
HD5980 & 5980 & WN4+O7I & 19.3 & 11.31 & 0.04 & 5100 & 5100\\
V444 Cygni & 193576 & WN5+O6II-V & 4.21 & 8 & 0.02 & 11100 & 11000\\
WR69 & 136488 & WC9d+OB & 2.293 &9.43& 0.015 &15600&13680\\
WR127 & 186943 & WN3b+O9.5V & 9.555 & 10.33 & 0.012&21400&17160\\
WR21 & 90657 & WN5o+O4-6 & 8.25443 & 9,76 & 0.011&22800&18720\\
&&&&&&&\\
\hline

\end{tabular}

\noindent *UV Flux at 1550A ($10^{-10}$ erg/s/cm$^2$/\AA)\\

\noindent $\dagger$ at 1500\AA, D=60 cm,  R=30000/N for spectroscopy
    \label{tab:target_list}
    
\end{table*}

\subsection{Spectroscopic Experimental Design}

In order to monitor the spectroscopic variability caused by selective wind eclipses and the presence of a colliding wind shock cone and to capture all the subtleties of the phase-related changes, we will use Polstar to secure a series of 10 high-resolution spectra, well-distributed over the orbital phase of each binary on our target list. We plan to observe the 8 brightest systems at the highest resolution (R=30000), the 6 next bright binaries at a slightly lower resolution (R=20000) and finally the 6 less UV bright targets at a more moderate resolution (R=10000). Including fainter targets broadens our system parameters and allows
us to include especially interesting targets with overlap with other \textit{Polstar} objectives,
such as V444 Cygni, a well-studied WR eclipsing binary, HD5980, for which one component star has recently undergone an LBV-type eruption, WR 69, a system including a WC9 type star that produces dust, and WR 137,
a rare system consisting of a WR star and an Oe star (with a decretion disk) in a wide orbit.

For the three brightest targets, we require a signal-to-noise ratio (SNR) of 3000 to reach the necessary
high polarimetric precision, while also achieving the
high spectral resolution needed to detect the details of the selective wind absorption profile, and details of the clumpy nature of the winds and of the colliding-wind shock-cone. For the other systems, we aim for a SNR of 100 (already 5-6 times higher that what is currently available from IUE data), which will still provide excellent quality data. We plan to use the knowledge gained with our test systems to enhance the interpretation of the lower SNR observations.
\
\ 
\subsection{Polarimetric Experimental Design}

We have selected three massive binaries that are bright in the ultraviolet to observe at the very highest spectroscopic resolution in spectropolarimetry to serve as test cases. These include hot, massive stars having reached different stages of evolution from bright giants to WR type. 
Although our standard polarization precision for this experiment is
$1 \times 10^{-3}$, for these unusually bright systems, 
we aim for a SNR of 3000 in order to be able to detect the very faint polarization signals from the colliding wind shock-cone,
down to a precision of $5 \times 10^{-4}$. 
We also plan to observe these targets in channel 2 at lower spectral resolution, and appropriately shorter
exposures, in order to characterize the wavelength dependency of the polarization at similar precision. 
For the other binaries, our strategy is to obtain the highest SNR possible  for the UV flux of the target (either 3000 or 1000) but at lower spectral resolution using observations in channel 2. Channel 1 observations at a lower SNR are also important to characterize the wavelength dependency of the polarization.  For all polarization measurements of our targets, we will be able to remove the non time-variable but wavelength-dependent interstellar polarization vector from the interstellar medium between each target and the observer, using the method described elsewhere in this topical collection by Andersson et al.
\
\vskip 1.5truecm

\section{Polarisation modeling}\label{sec:pol}

\subsection{Linear Polarization Across Spectral Lines}

At optical and UV wavelengths it is often a good approximation to assume that electron scattering is coherent in the frame of the electron. However, since the electron is moving (due to both thermal and large scale motions) relative to both the source and observer there will be a wavelength shift in the
observer's frame \citep{Auer1972}. Thermal motions can lead to either a decrease or increase in the scattered photon's wavelength, while scattering by an expanding  monotonic flow will always lead to a redshift.
These velocity shifts are observed  -- in P~Cygni for example, in which the thermal electron velocities are
larger than the wind velocity, the Balmer series, among others, show nearly
symmetric broad extended wings centered on the emission profile \citep{Bernat1978}. By contrast, in the spectra of many WR stars only a red electron scattering wing is seen, since in these stars the wind velocities are larger than the electron thermal velocities \citep{Hillier1984}.

\begin{figure*}[t]
\centering
\includegraphics[width=10 cm]{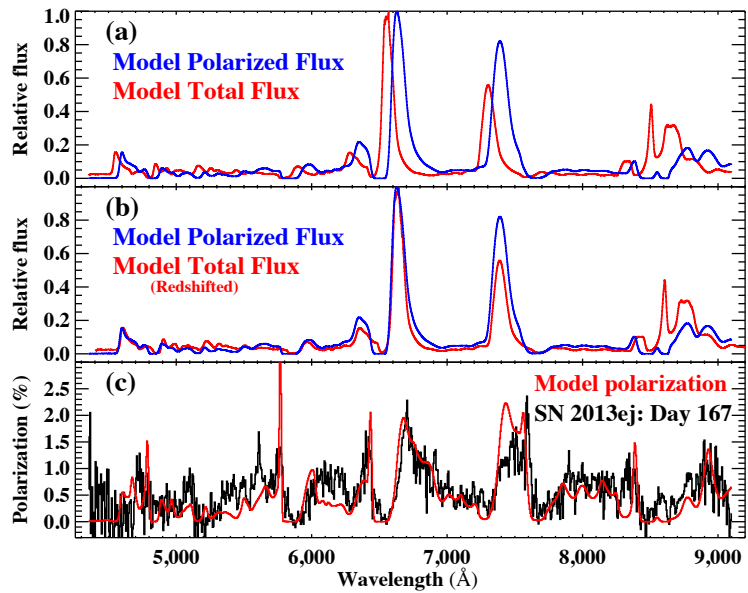}
\caption{Illustration of how velocity shifts influence the polarized spectrum. Top panel shows how the scattered spectrum (blue) is offset to the red from the flux spectrum (red). The second panel corrects for the offset, illustrating the similarities between the two spectra. However, because spectral features arise over a range of radii, the agreement is not perfect. The bottom panel shows the good agreement between a model (red) and observation (black) for SN 2013ej.  Figure from \cite{Leonard2021}. Reproduced with permission.}
\label{Fig_SN_pol}
\end{figure*}

For continuum polarization the velocity shifts induced by electron scattering are generally unimportant. However, they are of crucial importance in understanding line polarization. In Fig.~\ref{Fig_SN_pol} we show an extreme
example, taken from polarization studies of supernovae, where the line shift is crucial. In the Type II 
SN 2013ej the polarized spectrum is very similar to the observed spectrum \citep{Leonard2021}. However, it is different in two important ways: First,  the spectrum is redshifted. Second, the agreement between the intrinsic and polarized spectrum depends on which spectral feature is being examined.   Both effects are easily understood by assuming that the scattering source is offset from the SN emitting region, and by noting that
line emission in a SN region is stratified (different lines originate in different regions of the SN ejecta).  \cite{Dessart2021} argued that the scattering region was due to a nickel bubble, with an enhanced electron density
offset from the center of the ejecta. Due to motion relative to the source, the
spectrum is shifted to the red. Furthermore, the similarity between the intrinsic and scattered light will generally be 
highest for lines arising in the interior regions of the SN ejecta, since in that case the scattering angle is the same, unlike the case where the emission is formed over a much larger volume.

When optically thin electron-scattering is assumed, and other absorption processes neglected (e.g., purely bound-bound transitions), 
models predict that for an oblate spheroid the polarization should be parallel to the major axis. The reason is simple -- more photons will be scattered from the equatorial regions (with their
electron vector parallel to the major axis) than the polar regions. However, when optical depth effects are allowed for, the polarization can flip sign, as scattered photons become 
more isotropic in the higher optical depth equatorial region.

\begin{figure}
\includegraphics[width=8 cm]{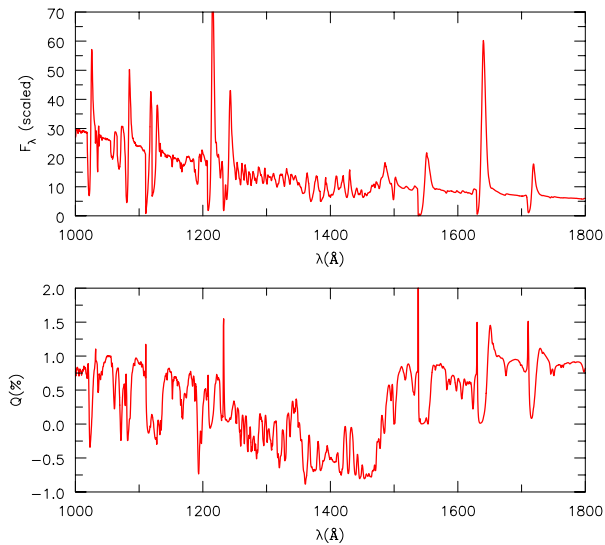}
\caption{The spectrum and polarization (Stokes Q) for a WR star with an oblate stellar wind viewed edge on. The ratio of the polar to equatorial density is 0.3. Notice how the polarization switches sign -- a consequence of optical depth effects in the wind.}
\label{Fig_WN_pol}
\end{figure}

The above scenario is illustrated in Fig. \ref{Fig_WN_pol} where we present the spectrum for a WN Wolf-Rayet star assuming it has an oblate wind with a polar to equatorial density ratio of 0.3. Near 1800\,\AA, the polarization is positive, as expected for an oblate spheroid. However, around 1400\,\AA, the polarization is negative, a consequence
of Fe-blanketing, which increases the total opacity and also introduces line effects.
The assumed axial symmetry guarantees that the position angle must be parallel or perpendicular to the rotation axis, so here we have taken the 
Stokes $Q$ component to subtract light polarized perpendicular
to the rotation axis from light polarized along the axis (and the $U$ component 
at 45 degrees to that would be zero in this symmetry).

Both of the above examples illustrate how the polarization contains important information about the structure of
circumstellar matter, whether it be a SN ejecta or a wind. Importantly, we now have the numerical tools to
interpret these observations. Polarization, in conjunction with other observations provides an invaluable tool to
help understand astrophysical objects, especially since most are spatially unresolved with modern telescopes. 

\subsection{Polarization from Colliding Wind Binaries}

While the preceding section focused on the spectropolarimetric signatures that can arise from aspherical winds of single stars, here the focus is on binary stars.  Whereas \citet[][]{PetersTC} emphasize interacting binaries, we explore variable wavelength-dependent polarization from massive stars with colliding winds.  The time dependence and wavelength dependence encode information about the geometry of the orbit (including viewing inclination that is important for obtaining accurate masses), the geometry of the bow shock, and properties of the respective stellar winds such as mass-loss rates.

\subsubsection{Continuum Linear Polarization of Colliding Wind Systems}

BME present a theoretical construction for thin electron scattering in
generalized envelopes with an arbitrary
number of illuminating point sources.  The two main limitations of BME are in relation to finite star effects (e.g., occultation) and the explicit assumption of thin scattering.  For application to binary stars, the former is often an adequate assumption unless the binary separation is quite close or the system is eclipsing.  For the latter assumption, the Thomson scattering cross-section is relatively small, and so thin scattering is often appropriate.

Importantly, despite the fact that Thomson scattering is a gray opacity,
a flat polarization
is not generally expected, particularly for a binary system.  
First, there can be changes across line features.
The influence of spectral lines come in 3 basic types:  (a) photospheric
lines, (b) recombination lines, and (c) scattering lines.  For
the scattering of starlight by electrons in a circumstellar envelope,
the polarization is flat across photospheric lines since the relative polarization is constant.  The line formation resides at the source, not in the extended circumstellar envelope.

For case (b), the recombination line is taken to form in the circumstellar
envelope.  It is possible for such line photons to be scattered by the circumstellar electrons.  But such photons are not originating from the star
but distributed throughout the envelope, and represent a diffuse source
of photons for scattering.  Variations in polarization can be complex.
As an example, \citet{Ignace2000} explored variations in polarization at H$\alpha$
in a Be star disk with a 1-armed spiral density wave:  the polarization in the continuum is little affected by the antisymmetric structure, but in H$\alpha$ the density-squared emissivity leads to a highly non-symmetric radiation field
in the disk.  The result is distinct polarimetric behavior within the line as compared to the continuum.  Similar consequences may apply to colliding wind systems \citep{Fullard2020thesis}.

There is an important limiting scenario for case (b) known as the ``line
effect'' \citep[e.g.,][]{1990ApJ...365L..19S}.  This is when the recombination line is formed at a large radius and the electron scattering occurs mainly over a more compact region close to the star.  In this situation the line photons are little scattered, and strong line emission can contribute to the flux $F_I$ but not $F_Q$ or $F_U$.  The result is a reduction in the relative polarization $p$ (sometimes referred to as ``diluted'' polarization).

Case (c) refers to the possibility of scattering resonance lines as an additional polarigenic opacity.  Although resonance scattering polarization for stellar winds has been explored in limited applications \citep{1989ApJS...71..951J, 2000A&A...363.1106I}, the topic has lacked an observational driver.  Such effects require a high-resolution UV spectropolarimeter and the computational tools for making predictions of effects and providing fits to data.  The data have largely not existed, given that {\em WUPPE} was of lower spectral resolution.
{\em Polstar} will provide new possibilities for case (c), and the team possesses the radiative transfer tools for developing diagnostics that combine both electron and resonance line scattering.

For hot stars in addition to the effects across lines are the chromatic effects in the relative polarization that arise simply from the presence of two stars with different temperatures in a binary.  The net polarization from the unresolved binary systems involves a weighted linear sum of the polarized contributions from scattered starlight by each star.  The net polarization is of the form:

\begin{equation}
    p(\lambda) = \epsilon_1p_1+\epsilon_2p_2,
\end{equation}

\noindent where 

\begin{equation}
   \epsilon_1 = \frac{L_1(\lambda)}{L_1(\lambda)+L_2(\lambda)}   , 
\end{equation}

\noindent and

\begin{equation}
   \epsilon_2 = \frac{L_2(\lambda)}{L_1(\lambda)+L_2(\lambda)}    ,
\end{equation}

\noindent and $p_1$ and $p_2$ represent the relative polarization caused strictly by geometric effects from the perspective of each separate star.  These two parameters bracket the overall scale of the net polarization for an unresolved source.  Neither are functions of wavelength; chromatic effects in the polarization arise from the luminosity weighting by the two stars.

To understand general expectations for $p(\lambda)$, it is helpful to consider limiting cases.
If the temperatures of the two binary components are identical, the polarized spectrum will be flat at all wavelengths, in the absence of multiple scattering effects. If the stars have different temperatures, then the polarized spectrum, $p(\lambda)$ will not in general be flat.  In particular, for hot massive stars, the deviation from a flat polarized spectrum occurs mainly in the UV.  Thinking simplistically in terms of Planckian stellar spectra, hot OB star spectra in the optical and longer wavelengths will be in the Rayleigh-Jeans regime.  At such long wavelengths, $p(\lambda)$ will be flat because both $L_1$ and $L_2$ have the same spectral shapes.  It will be flat until bound-free or free-free emission becomes relevant.  In particular, a WR star does not achieve the Rayleigh-Jeans limit at any wavelength owing to significant
 free-free and bound-free emission from the wind.  By contrast at short wavelengths in the UV, where {\em Polstar} will operate, the presence of differing Wien peaks ensures that $\epsilon_1$ and $\epsilon_2$ vary with wavelength, and $p(\lambda)$ will be chromatic.  Formally, $p(\lambda)$ is bounded by $p_1$ and $p_2$, which provides a richer set of constraints for inferring the geometry of the system. 

For our application to massive colliding wind binary systems, we assume that the colliding wind interface (CWI) is axisymmetric about the line-of-centers joining the two stars, following \cite{Canto1996}.  We further assume that the separate winds of the two stars are each spherically symmetric up to the CWI.  
The {\em Polstar} team possesses a suite of computational tools for time-dependent hydrodynamical simulation of colliding winds \citep[e.g.,][]{2017MNRAS.464.4958R} and for the radiation transport for synthetic polarization spectra \citep{Hoffman2007,Huk2017}.  Here we explore only the BME approach as the underlying assumptions can often be valid or nearly so.    

\subsubsection{Model Prescription}\label{sec:canto}

\cite{Canto1996} present a convenient semi-analytic bow shock model solution for the shock properties resulting from the collision of two spherical winds at terminal speed under the condition of strong radiative cooling\footnote{Some colliding wind shocks are better described by adiabatic cooling \citep[e.g.,][]{2009ApJ...703...89G}.  For our purposes the solution of \cite{Canto1996} is convenient for its simplicity and for overall characterization of the issues involved for polarimetric variability.}.  In this model, the bow shock geometry and its surface density is determined by two fundamental ratios:

\begin{eqnarray}
\beta & = & \frac{\dot{M}_2v_2}{\dot{M}_1v_1},~{\rm and} \\
\alpha & = & v_2/v_1,
\end{eqnarray}

\noindent where ``1'' and ``2'' refer to the two stars, $v$ is the wind terminal speed, and $\dot{M}$ is the mass-loss rate.  We identify the primary as star \#1 in such a way that the ratio of wind momenta is $\beta\le 1$.   By this convention the ratio of terminal speeds, $\alpha$, may be smaller, larger, or equal to unity.  Note also that the primary by our definition may not be the most luminous star in the system for all (or any) wavelengths.  The key outcomes of their solution is the location of the bowshock and its geometry, along with the surface density of material in the bow shock.

\begin{figure}
\centering
\includegraphics[width=\columnwidth]{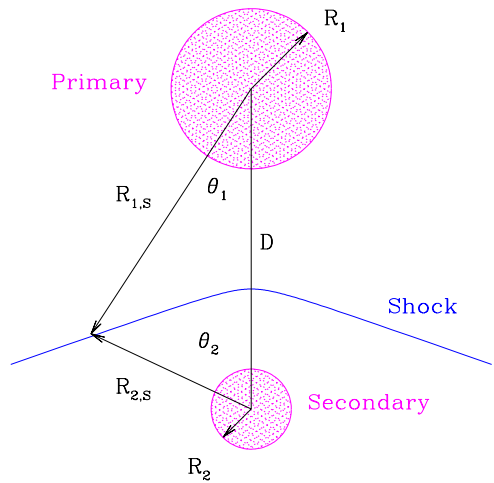}
\caption{Top-down illustration of the two stars (primary as ``1'' and secondary as ``2'') and the bowshock formed by the colliding winds.  Labeled variables are defined in text.  \citep[Figure similar to one in][]{Ignace2022}
\label{fig:bowshock}}
\end{figure}

The individual winds are taken
to be spherical at terminal speed with densities varying with the inverse square of the distance from the star.  The semi-analytic
solution of \cite{Canto1996} yields the geometry of the bowshock in terms of distances, $R$, from the secondary and the primary stars as (refer to Fig.~\ref{fig:bowshock} for a schematic): 

\begin{eqnarray}
R_{2,S} = D\,\frac{\sin \theta_1}{\sin(\theta_1+\theta_2)},~{\rm and} \\
R_{1,S} = \sqrt{D^2+R^2_{2,S}-2\,D\,R_{2,S}\,\cos\theta_1},
\end{eqnarray}

\noindent where $D$ is the separation between the two stars at any moment.
Naturally the solution is only valid when $D>R_{2,\ast}$.
A special quantity is the standoff radius of the bowshock along
the line centers, denoted as $R_{1,0}$ and $R_{2,0}$, given by

\begin{eqnarray}
R_{2,0} = \frac{\beta^{1/2}}{1+\beta^{1/2}}\,D,~{\rm and} \\
R_{1,0} = \frac{1}{1+\beta^{1/2}}\,D.
\end{eqnarray}

\noindent The case $\beta=1$ corresponds to a planar shock between
identical stars and winds, with $\theta_{1,\infty}=\theta_{2,\infty}=\pi/2$.
The parameters $p_1$ and $p_2$ are set by the binary orbital separation $D$, the wind
properties, and the viewing inclination.  For a circular orbit, $D$ is fixed and so
are $p_1$ and $p_2$.  Yet there can be time dependence of the polarization and position
angle with orbital phase that arises as the stars orbit each other (c.f., BME).
If the orbit is eccentric, then separation $D$ is a function of orbital phase
as well.  The calculation of $p_1$ and $p_2$ is detailed in \citet{Ignace2022}.

\subsubsection{Model Results}

For illustration, we use the approach of \citet{Ignace2022}, and evaluate synthetic spectra with linear polarization for a binary involving a WR star and an O star.  For SEDs we use the online PoWR model grid\footnote{www.astro.physik.uni-potsdam.de/~wrh/PoWR/powrgrid1.php} \citep{2002A&A...387..244G, 2015A&A...577A..13S, 2019A&A...621A..85H}.  For the WR spectrum, we selected model 08-10 from the Milky Way WNE models.  For the O star, we used model 36-40 at Galactic metallicity.  Table~\ref{tab:binaryparams} lists the stellar and wind parameters.

\begin{table}\caption{Stellar and Wind Parameters for Colliding Wind Polarization}
    \centering
    \begin{tabular}{ccc}
\hline      & WN5/6 & O7V \\ \hline
 $T$ (kK)     &  56.2 & 36.0 \\
 $\log L/L_\odot$      & 5.30  & 5.01 \\
 $\log \dot{M}$ $(M_\odot$ yr$^{-1}$)     & $-5.13$  & $-7.00$ \\ 
 $v_\infty$ (km/s)    & 1600  & 2710 \\ 
  $R_\ast/R_\odot$ & 4.7 & 8.2 \\
  $M_\ast/M_\odot$    &  25  & 25 \\ \hline
    \end{tabular}
    \label{tab:binaryparams}
\end{table}

\begin{figure}
\centering
\includegraphics[width=\columnwidth]{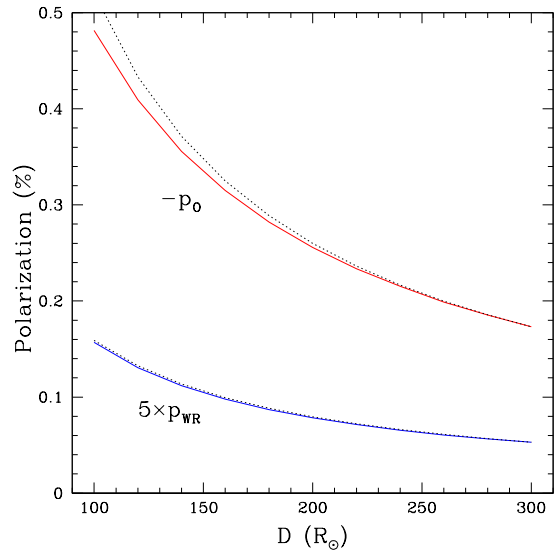}
\caption{Plot of $p_{WR}=p_1$ and $p_O=p_2$ for a WR+O binary using the \citet{Canto1996} bow shock solution against binary separation $D$ in $R_\odot$.  Stellar and wind parameters are given in the text.  With $p_O<0$, the negative of its value is plotted (red).  With $p_{WR} \ll -p_O$, the value has been increased by $5\times$ for ease of viewing (blue).  The dotted lines are $D^{-1}$ curves showing that $p_{WR}$ closely follows this trend for all separations, whereas $p_O$ approaches it for wider separations.
\label{fig:polparams}}
\end{figure}

Figure~\ref{fig:polparams} shows values of $p_1 = p_{WR}$ (blue) and $p_2=p_O$ (red) in percentage for these example stellar and wind parameters plotted with binary separation $D$ in solar radii.  The parameter $p_O<0$, so its negative is plotted.  For the WR star, its wind is far stronger than the O star, so the bow shock is located far from the WR component and close to the O component.  As a result, $p_{WR} \ll -p_O$ and so it is plotted at $5\times$ its value for ease of viewing.  Also shown are dotted lines that scale as $D^{-1}$ for binary separation.  The parameter $p_{WR}$ closely follows that trend, whereas $p_O$ follows it only at larger separations.  But even at relatively close separations, the trend of $p_O$ is close to $D^{-1}$.

\begin{figure}
\centering
\includegraphics[width=\columnwidth]{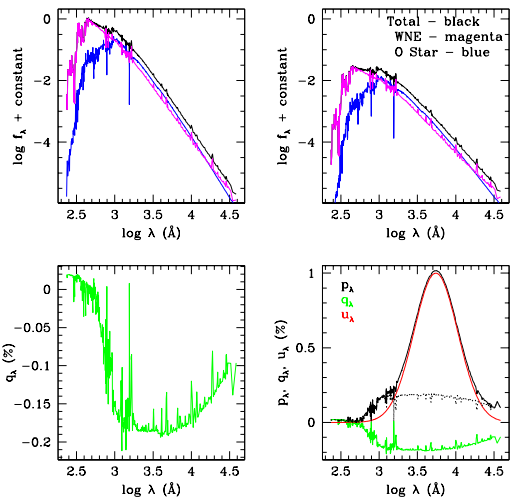}
\caption{(Upper left) The spectral energy distributions of the WR star (magenta), O star (blue), and combined (black).  (Upper right) The spectra from the upper left panel now with interstellar extinction included as described in the text.  (Lower left) The intrinsic polarization spectrum from the binary entirely in $q_\lambda$, as described in the text.  (Lower right) The polarized spectrum now including a Serkowski Law as described in the text.  Here $q_\lambda$ in green is entirely stellar; $u_\lambda$ in red is entirely interstellar; $p_\lambda$ in black is the total polarization; and the dotted black line is $p_\lambda =  \sqrt{ q_\lambda ^2}$ if there had been no interstellar
polarization.
\label{fig:polspec}}
\end{figure}

The PoWR model spectra are plotted in the upper left panel of Figure~\ref{fig:polspec}, with magenta for the WR component, blue for the O component, and black for the summed spectrum. Note that the total spectrum is normalized to have unity peak flux.  The upper right panel includes interstellar extinction at a value of $A_V=2$.  Ignoring the ``blue bump'' \citep{1989ApJ...345..245C}, we use an extinction law of $A_\lambda = A_V (0.5 - 1.07\lambda_\mu)$, where the wavelength is in microns.  While ignoring the blue bump fails to properly characterize the expected observed spectrum, it does not affect the effect of interstellar polarization described by the Serkowski Law \citep{1988ApJ...327..911C, 1999ApJ...510..905M}, which displays no blue bump features.

At lower left of Figure~\ref{fig:polspec} is the polarized spectrum without interstellar polarization using $p_{WR} = + 0.0202$ and $p_0 = - 0.32$ for a binary separation of $D=160R_\odot$ from
Figure~\ref{fig:polparams}.  Note that for the masses of Table~\ref{tab:binaryparams}, the orbital period for this illustrative example would be about 45 days. We have oriented the system as pole-on and with all the polarization in Stokes-Q.

There are several key features regarding this figure.  First, whenever the WR flux dominates, the polarization is positive and close to zero.  This is evident around EUV wavelengths.  Whenever the polarization is significantly negative, the O star tends to dominate.  This occurs from the FUV through the optical.  For IR wavelengths, the net polarization reduces as the free-free continuum of the WR component gradually gains in brightness over the O star.  Also notable is that the polarization may become larger or smaller across lines, depending on whether a strong line belongs to the WR or O star component.

At lower right is the stellar polarization with interstellar polarization from a Serkowski Law now included.  The Serkowski Law is given by

\begin{equation}
    p_{ISM}(\lambda) = p_{\rm max} \,\exp\left[-K\,\ln(\lambda/ \lambda_{\rm max})^2\right],
\end{equation}

\noindent where $p_{\rm max}$ is the peak polarization for which we choose 1\%, $K$ has a typical value of 1.15, and $\lambda_{\rm max}$ is the wavelength of the peak interstellar polarization, which has a typical value of 550 nm.  
We orient the polarization at a position angle so it is entirely in Stokes-U.  The lower right panel shows $q_\lambda$ in green (same as shown in the lower left panel), which is entirely stellar, $u_\lambda$ which is entirely interstellar, $p_\lambda=\sqrt{q_\lambda^2+u_\lambda^2}$ in black, 
and then $q_\lambda$ as the dotted black line.  Interstellar polarization strongly affects the polarization in the optical band, whereas the stellar polarization becomes dominant toward UV and IR wavelengths.

\begin{figure}
\centering
\includegraphics[width=\columnwidth]{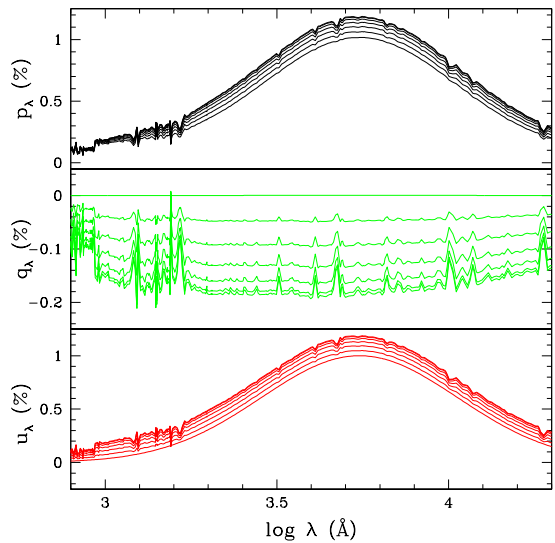}
\caption{The three panels display total polarization (top), Stokes-Q (middle), and Stokes-U (bottom) for half a binary orbit as seen pole-on.  Interstellar polarization is entirely in $u_\lambda$, whereas the stellar polarization smoothly varies between $q_\lambda$ and $u_\lambda$ owing to the changing orientation on the sky.
\label{fig:polorb}}
\end{figure}

Figure~\ref{fig:polorb} displays the results of Figure~\ref{fig:polspec} for different polarization position angles assuming a circular orbit and a pole-on view.  The upper panel is the total polarization; middle is for Stokes-Q; and bottom is for Stokes-U.  Note that the scale for the lower panel differs from the other two.
With our arrangement of the interstellar polarization position angle, the modulation of $q_\lambda$ with orbital phase is entirely stellar.  By contrast, $u_\lambda$ is a blend of stellar and interstellar.  The variations are displayed for only half the orbit for which $q_\lambda\le 0$.    

For analysis of observational data, it is important to note that the polarization position angle for the interstellar component is fixed and unvarying with time.  Thus, if a binary system were strictly axisymmetric, there would be a position angle for which the Stokes $q_\lambda$ and $u_\lambda$ could be rotated to eliminate (through fitting) the Serkowski Law from infuencing one of these Stokes parameters, such as $u_\lambda$ in our illustrative example.  In this case the influence of interstellar polarization would be relegated entirely to $q_\lambda$. 
However, if the stellar system is not strictly axisymmetric, then there is no rotation of the observed Stokes-Q and U components in which the stellar polarization can be isolated to a single Stokes parameter, with the interstellar contribution limited to the other.  The interstellar polarization can still be eliminated through analysis of the $q_\lambda-u_\lambda$ variations with time, but in a model-dependent way. 

\subsection{Numerical models}

Radiative transfer models are capable of modeling more complex binary phenomena than those allowed by the assumptions of BME \citep[e.g., ][]{Kurosawa2002,Hoffman2003}. Leveraging these models is essential for any deep investigation of specific colliding wind systems. Polarization of spectral lines is especially suited to numerical modeling because of how difficult the problem is in asymmetric geometries. 

\begin{figure}
\centering
\includegraphics[width=0.9\columnwidth]{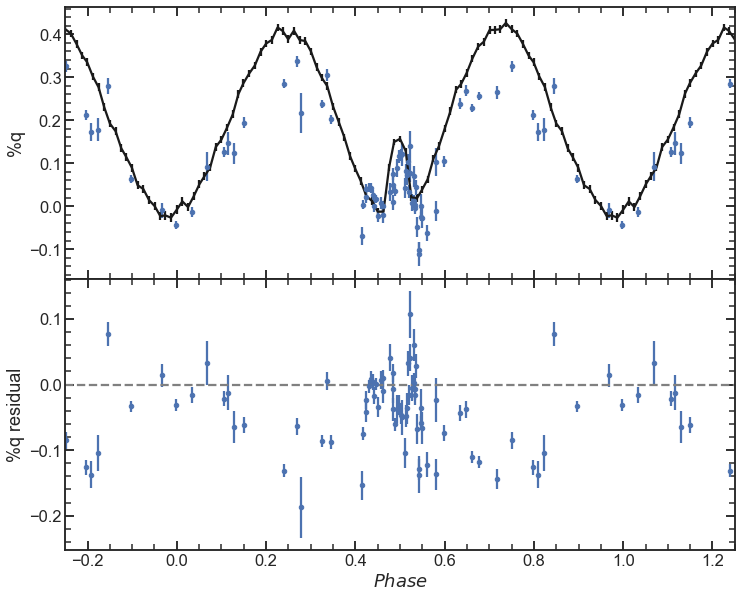}
\includegraphics[width=0.9\columnwidth]{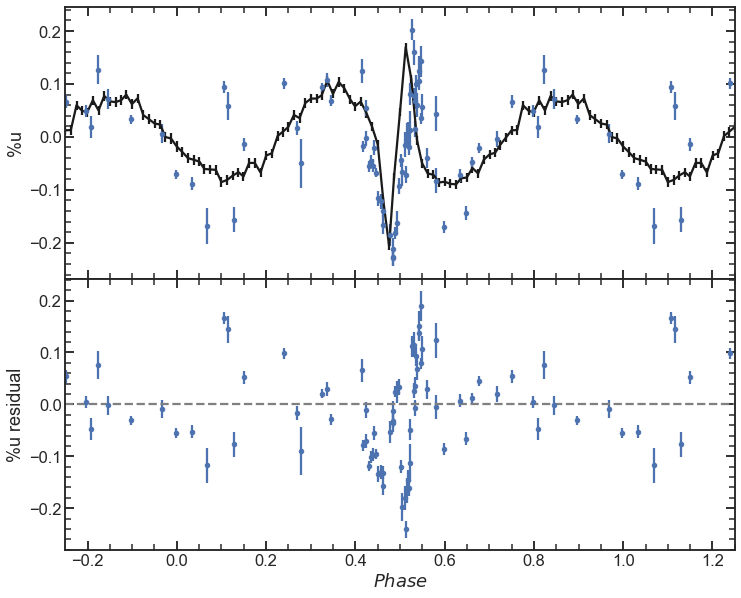}
\caption{V444 Cygni polarization (blue points) compared to the \citet{Kurosawa2002} E model at an inclination angle of 82.2$^{\circ}$ (black line), with  rotated O star wind cavity as in the emission line model. The $R$-band polarization data are taken from \citet{St.-Louis1993}. Residuals are also presented below each plot. Reproduced with permission.}
\label{fig:v444_toy_continuum}
\end{figure}

Numerical models can also be used as initial probes much like BME. Because electron scattering is the primary polarigenic mechanism in hot star winds, the effects of line and continuum polarization can be decoupled by considering their emission and scattering regions. A toy model of the well-studied V444 Cygni has been constructed for this purpose \citep{Fullard2020thesis}. It is based on the \citet[][]{Kurosawa2002} E model of the system including a WR star with an appropriate wind density profile from the Potsdam Wolf-Rayet (PoWR) models\footnote{See the Potsdam PoWR models at the following website: www.astro.physik.uni-potsdam.de/~wrh/PoWR/powrgrid1.php}. A simple spherical cap cutout represents the O star wind (assumed to be of negligible relative density) at the wind collision point, rotated to mimic the effect of orbital motion. Emission occurs from the WR and O star photospheres (where the WR star photosphere is defined from its PoWR model). It reproduces the continuum behaviour of the system as shown in Figure~\ref{fig:v444_toy_continuum}. 
 
\begin{figure}
\centering
\includegraphics[width=\columnwidth]{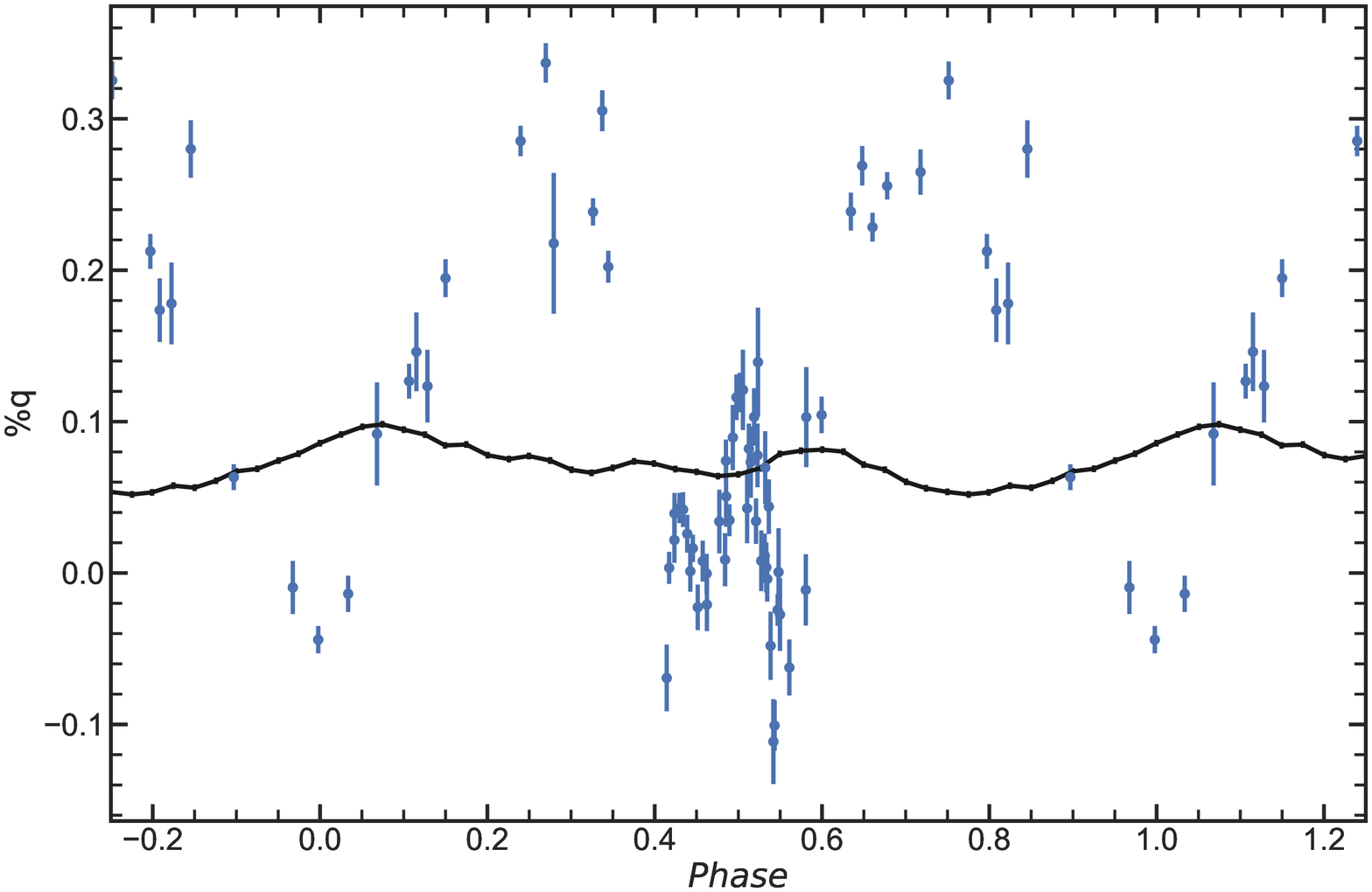}
\includegraphics[width=\columnwidth]{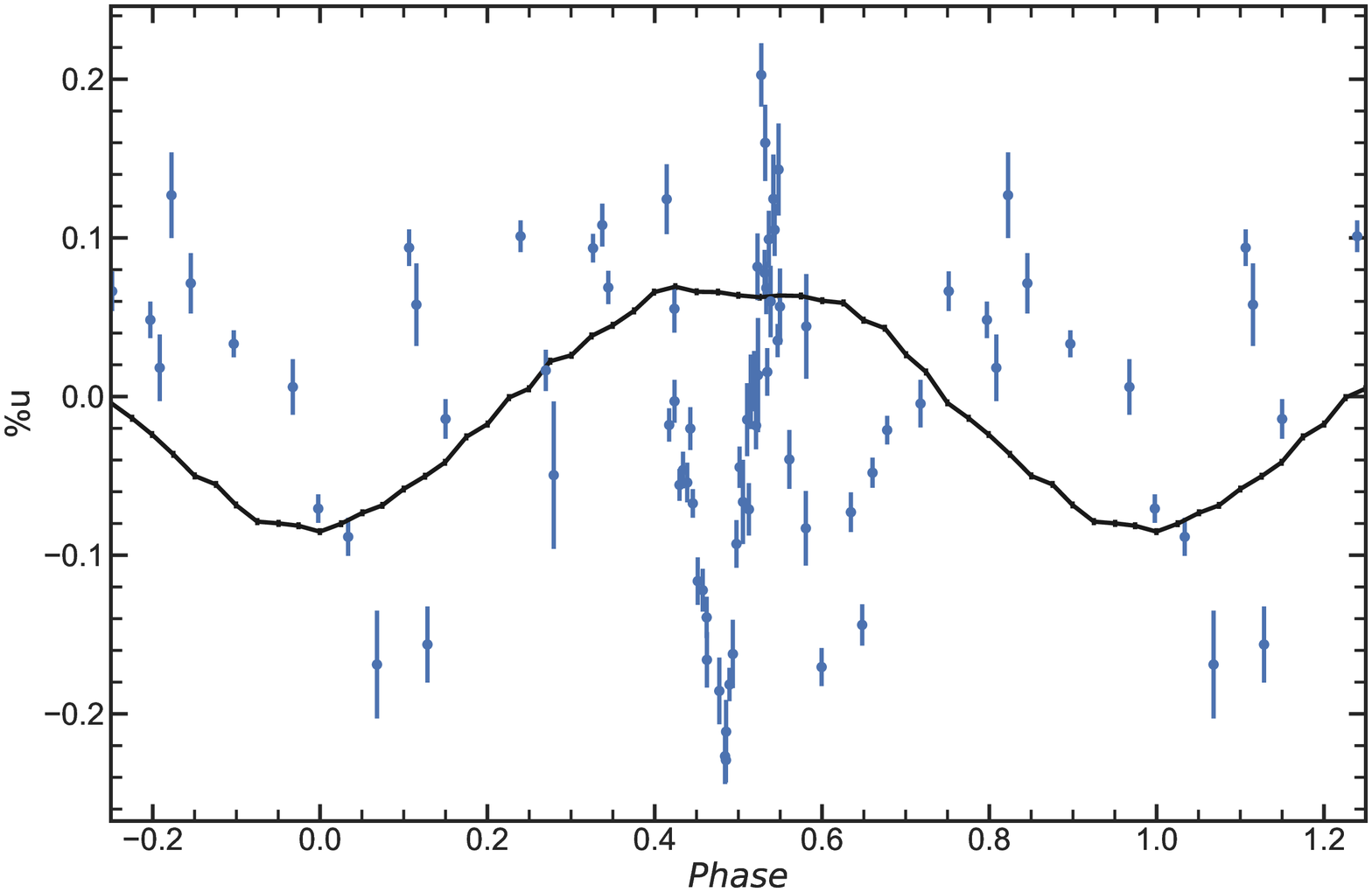}
\caption{V444 Cygni polarization (blue points) compared to a wind + shock emission line region model at an inclination angle of 82.2$^{\circ}$ (black line). The $R$-band polarization data are taken from \cite{St.-Louis1993}. The comparison between the expected signal in a He{\sc ii} line from a wind and CWI and the observations in continuum light clearly shows that the signatures are completely different but are both of similar amplitude (in this case only in u). Reproduced with permission.}
\label{fig:stlouis_windshock_slip_compare}
\end{figure}

\begin{figure}
\centering
\includegraphics[width=\columnwidth]{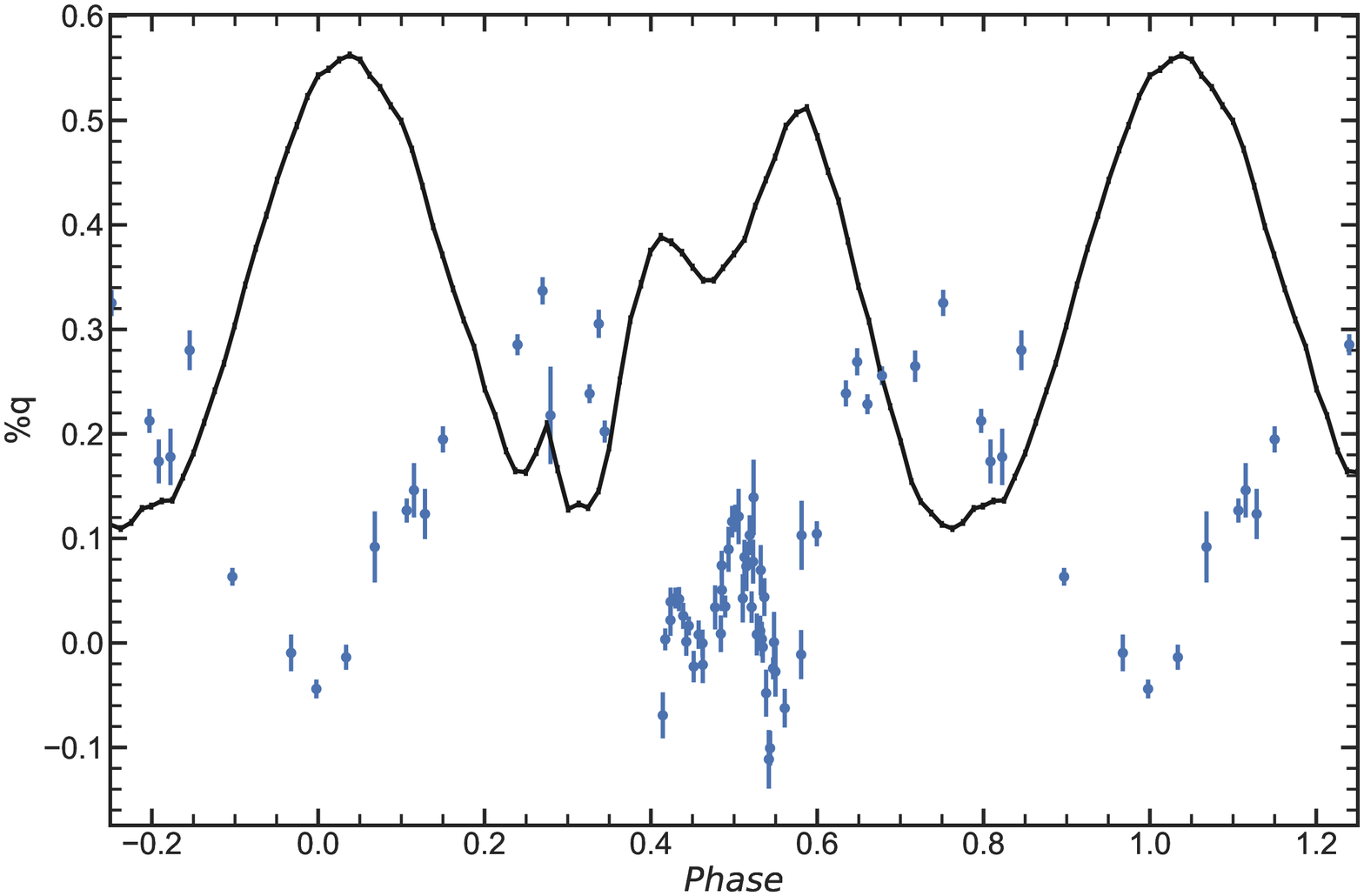}
\includegraphics[width=\columnwidth]{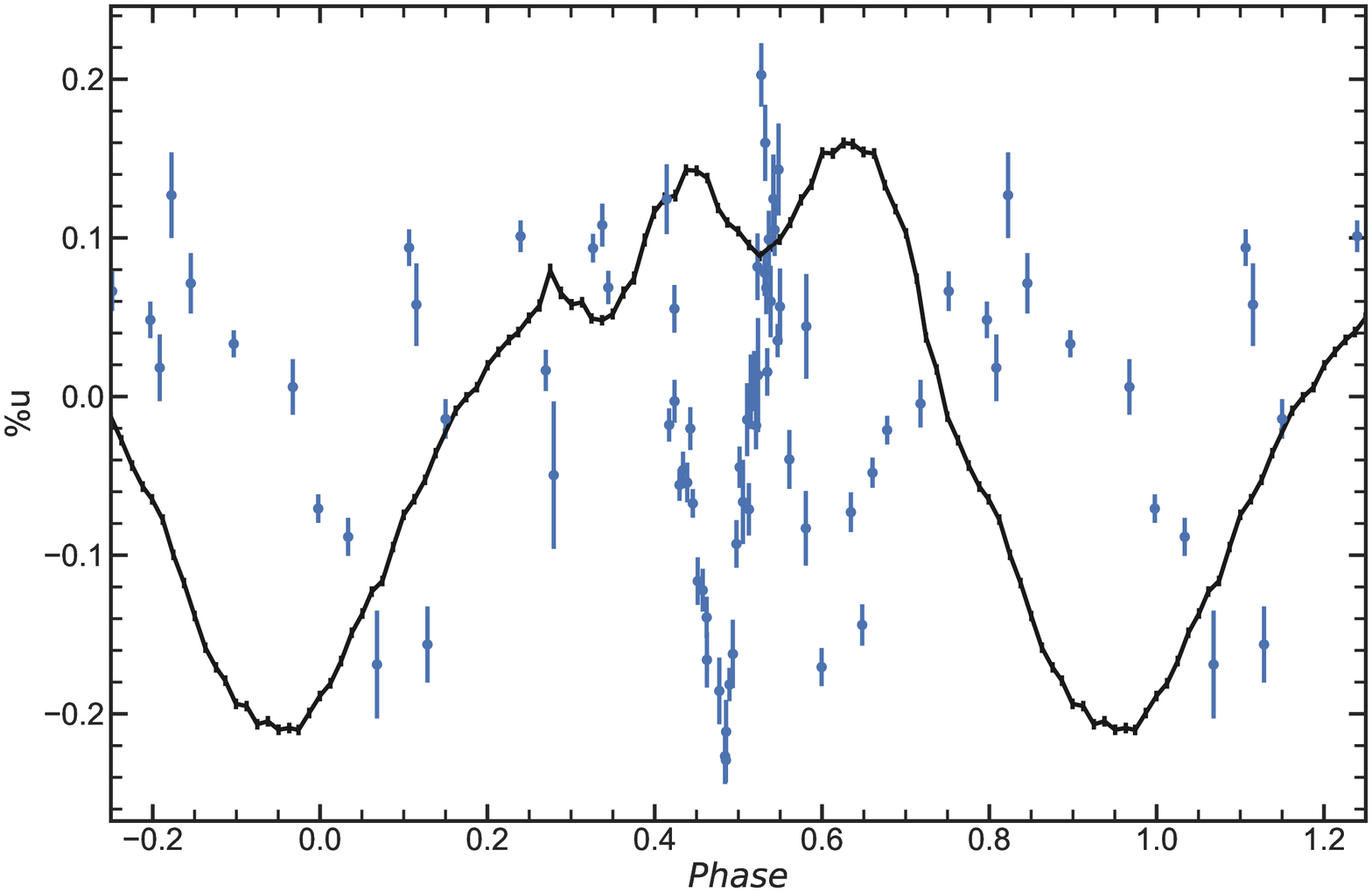}
\caption{V444 Cygni polarization (blue points) compared to a shock emission line region model at an inclination angle of 82.2$^{\circ}$ (black line). The $R$-band polarization data are taken from \cite{St.-Louis1993}.The comparison between the expected signal in a He{\sc ii} line from a CWI and the observations in continuum light clearly shows that the signatures are completely different but are both of similar amplitude. Reproduced with permission.}
\label{fig:stlouis_shock_slip_compare}
\end{figure}

To represent line polarization caused by He{\sc ii}, the emission locations are simply changed to originate from a shell in the WR star wind, and from the edge of the shock region in two clumps. The clumps are located at phase $\sim0.12$ and $\sim0.75$, and lie at the edge of the simulation just beyond the O star orbit. Figures~\ref{fig:stlouis_windshock_slip_compare} and~\ref{fig:stlouis_shock_slip_compare} show the resulting polarization curves compared to existing continuum polarization data. Figure~\ref{fig:stlouis_windshock_slip_compare} shows the complete wind and shock model, which has a low near-constant Stokes $q$ and a similar amplitude of variation in Stokes $u$. Figure~\ref{fig:stlouis_shock_slip_compare} shows a model that produces emission only from the two shock regions. This has strong phase-dependent polarization in both Stokes $q$ and $u$. Neither model produces polarization similar to the continuum signal. Both show the importance of considering line polarization and its capability to diagnose emission regions in colliding winds. It is clear that obtaining polarization observations specifically in spectral lines will provide additional information than that of the continuum polarization.

\section{Conclusions}\label{sec3}
Colliding-wind binaries are important laboratories for the study of radiative wind driving, 
and massive stars in general, as the detailed structure and geometry of the interaction region between the winds can yield important information about each component's individual wind properties. This interaction is not only due to the bow shock, since the wind from each star can also be influenced by the radiation field from the other star. This phenomenon is best studied with a combination of spectroscopy and polarimetry, and as demonstrated above, the ultraviolet provides an ideal bandpass since the spectral energy distributions of massive stars peak in that region, leading to a wavelength dependence of the polarization. 

Several analytical models provide a satisfactory description of the wind collision region, within certain simplifying assumptions detailed in Sect.~\ref{sec:pol}. Our group possesses the ability to leverage these models, as demonstrated for a few illustrative cases.

Using time series of spectropolarimetric observations from Polstar, we will be able to constrain the geometry of the bow shock and therefore the wind properties of a sample of 20 colliding-wind binaries spanning a large parameter space. The obtained dataset will significantly improve upon available observations; spectroscopy will be obtained with much higher signal-to-noise ratio and resolution than with IUE and WUPPE, and with greater polarimetric precision than the latter. 
Achieving a SNR of 100 at spectral resolution down to the wind sound speed,
over an orbitally resolved time domain, opens up new and unique wind dynamics diagnostics, as
the light from one star scatters in, and is absorbed by, the wind of the other, with orbitally
modulated changes in the relevant angles.
These produce a more precise mapping between radius and velocity than possible
for single stars, allowing tests of the degree to which different wind types and optical depths
alter the wind acceleration and structure formation.
Information about the bow shock also constrains the relative momentum fluxes in the winds
of the two different stars in the binary.

The work described in this paper corresponds to a single science objective included
in the \textit{Polstar} proposal, which overlaps in interesting ways
with several other objectives described in their own individual papers.
We will take advantage of the ways 
the unique diagnostics supplied by colliding wind binaries informs these other objectives.
For example, \citet[][]{GayleyTC} discuss how \textit{Polstar} 
enables the study of structure and clumping in the winds of the $\sim 40$ brightest targets, in the context of single stars observed continuously over wind dynamical timescales.
Two of those targets, $\gamma$ Vel and $\delta$ Ori, are binary systems in the colliding-wind
target list, so our observations sampled over the system orbit will provide important
additional context for the continuous wind dynamics observations.
Also, \citet[][]{udDoulaTC} explore the global structural effects of
strong magnetization in a subset of OB stars, and a particularly important
example is Plaskett's star, which is the only strongly magnetic star observed to also be a rapid rotator.
This star is also in our colliding wind target list, and the understanding of
its magnetic structure will be complemented by the methods made available by its short-period binary status,
as described here.

Closer binaries than the ones examined in this paper can interact and undergo episodes of 
conservative and nonconservative mass transfer, altering the mass and 
angular momentum of the binary constituents in ways
that are the subject of other \textit{Polstar} objectives and described 
in their own papers. \citet[][]{PetersTC} investigate how UV spectropolarimetry can characterize these mass flows and help determine the fraction of the mass loss by the system, whereas \citet[][]{JonesTC} focus on angular momentum evolution and the pathways
for stellar spinup and Keplerian disk creation.  UV spectropolarimetry can also provide information regarding the geometry, velocity, and inclination of such disks, as well as the accretion disks of Herbig Ae/Be stars \citep{WisniewskiTC}.

All of these objectives, whose realization will be made possible by 
\textit{Polstar}, constrain key evolutionary stages of massive stars and lead to a better understanding of their fates, including the compact objects that they leave behind. Importantly, the analysis of polarization signals from all of these sources will require a detailed understanding of the polarization induced by the intervening interstellar medium; by contributing to improve this knowledge, Polstar will not only enable these topics of stellar physics, but will also yield important insights in the physics of interstellar dust grains \citep[][]{AnderssonTC}.
Any of these types of systems can, in special cases, provide examples that can be studied over an orbitally modulated range of angles that are either completely sampled, or
at least change considerably, over the timeframe of the \textit{Polstar} mission.
This paper, therefore, is focused on the complementary spectropolarimetric
diagnostics available in 20 such cases.

\section*{Affiliations}

$^{1}${\orgdiv{D\'epartement de physique, Universit\'e de Montr\'eal, Complexe des Sciences, 1375 Avenue Th\'er\`ese-Lavoie-Roux, Montr\'eal (Qc), H2V 0B3, Canada}}

$^{2}${\orgdiv{Department of Physics and Astronomy, University of Iowa, Iowa City, IA, 52242}}

$^{3}${\orgdiv{Department of Physics and Astronomy \& Pittsburgh Particle Physics, Astrophysics and Cosmology Center (PITT PACC)}}

$^{4}${\orgdiv{Department of Physics \& Astronomy,East Tennessee State University,Johnson City, TN 37614, USA}}

$^{5}${\orgdiv{Department of Physics and Astronomy, Western University, London, ON N6A 3K7, Canada}}

$^{6}${\orgdiv{Department of Physics and Astronomy, Howard University, Washington, DC 20059, USA}}

$^{7}${\orgdiv{Center for Research and Exploration in Space Science and Technology, and X-ray Astrophysics Laboratory, NASA/GSFC, Greenbelt, MD 20771, USA}}

$^{8}${\orgdiv{Department of Physics and Astronomy, Embry-Riddle Aeronautical University, 3700 Willow Creek Rd, Prescott, AZ, 86301, USA}}

$^{9}${\orgdiv{Armagh Observatory and Planetarium, College Hill, BT61 9DG Armagh, Northern Ireland, UK}}

$^{10}${\orgdiv{Department of Physics \& Astronomy, University of Southern California, Los Angeles, CA 90089, USA}}

$^{11}${\orgdiv{Department of Physics \& Astronomy, University of Denver, 2112 E. Wesley Ave., Denver, CO 80208, USA}}

$^{12}${\orgdiv{GAPHE, Universit\'e de Li\`ege, All\'ee du 6 Ao\^ut 19c (B5C), B-4000 Sart Tilman, Li\`ege, Belgium}}

$^{13}${\orgdiv{Department of Physics \& Astronomy, University of Auckland, 38 Princes Street, 1010, Auckland, New Zealand}}

$^{14}${\orgdiv{Anton Pannekoek Institute for Astronomy and Astrophysics, University of Amsterdam, 1090 GE Amsterdam, The Netherlands}}

$^{15}${\orgdiv{Department of Physics \& Astronomy, Michigan State University,567 Wilson Rd., East Lansing, MI 48824, MI, USA}}

$^{16}${\orgdiv{Physics Department, United States Naval Academy, 572C Holloway Rd, Annapolis, MD 21402, USA}}

$^{17}${\orgdiv{NASA GSFC, Greenbelt , MD 20771, USA}}

\bibliography{main}
\newpage
\section*{Ackowledgments \& Declarations}
\bmhead{Acknowledgments}

RI acknowledges funding support from a grant by the National Science Foundation (NSF), AST-2009412. JLH acknowledges support from NSF under award AST-1816944 and from the University of Denver via a 2021 PROF award.
Scowen acknowledges his financial support by the NASA Goddard Space Flight Center to formulate the mission proposal for Polstar.
Y.N. acknowledges support from the Fonds National de la Recherche Scientifique (Belgium), the European Space Agency (ESA) and the Belgian Federal Science Policy Office (BELSPO) in the framework of the PRODEX Programme (contracts linked to XMM-Newton and Gaia).
NSL and CEJ wish to thank the National Sciences and Engineering Council of Canada (NSERC) for financial support.
A.D.-U. is supported by NASA under award number 80GSFC21M0002.
GJP gratefully acknowledges support from NASA grant 80NSSC18K0919 and STScI grants HST-GO-15659.002 and HST-GO-15869.001.  
\section*{Declarations}

\subsection*{Funding}
The Polstar satellite is presently being proposed as a Midex mission to NASA and is currently under evaluation.
\subsection*{Author Contribution}
All authors have contributed ideas to motivate the work presented in this paper. Some have provided specific modelling, others proposed data analysis techniques. All contributed to the writing of the text to varying extents.
\subsection*{Competing Interests}
To our knowledge, there are no competing interests. 
\subsection*{Data Availability}
The team will make the data obtained within the context of this mission available to the astronomical community.

\end{document}